\documentstyle[onecolumn]{mn}

\input epsf.tex
\let\text=\rm
\begin{document}
\title{Coupled Quintessence}
\author[Luca Amendola]{Luca Amendola\\
Osservatorio Astronomico di Roma, \\
Viale Parco Mellini 84, \\
00136 Roma, Italy\\
{\it amendola@oarhp1.rm.astro.it}}
\date{Accepted...; Received... }
\maketitle

\begin{abstract}
A new component of the cosmic medium, a light scalar field or \
''quintessence '', has been proposed recently to explain cosmic acceleration
with a dynamical cosmological constant. Such a field is expected to be
coupled explicitely to ordinary matter, unless some unknown symmetry
prevents it. I investigate the cosmological consequences of such a coupled
quintessence (CQ) model, assuming an exponential potential and a linear
coupling. This model is conformally equivalent to  Brans-Dicke Lagrangians
with power-law potential. I evaluate the density perturbations on the cosmic microwave
background and on the galaxy distribution at the present and derive bounds
on the coupling constant from the comparison with observational data.

A novel feature of CQ is that during the matter dominated era the scalar
field has a finite and almost constant energy density. This epoch, denoted
as $\phi $MDE, is responsible of several differences with respect to
uncoupled quintessence: the multipole spectrum of the microwave background
is tilted at large angles, the acoustic peaks are shifted, their amplitude
is changed, and the present 8Mpc$/h$ density variance is diminished. The
present data constrain the dimensionless coupling constant to $|\beta |\leq
0.1$.
\end{abstract}

\section{Introduction}

\bigskip

The recent evidence in favour of an accelerated cosmic expansion (Perlmutter
et al. 1999; Riess et al. 1999) has prompted the theorists to hypothesize
components of the cosmic medium additional to the ordinary matter and
radiation, whose equation of state is unable to provide the required
kinematics. In a flat Universe, the dark energy of such a component should
provide roughly 70\% of the cosmic density, and should possess an effective
equation of state 
\begin{equation}
p=(w-1)\rho ,
\end{equation}
with the present value (Turner \& White 1997, Perlmutter et al. 1999, Wang
et al. 1999) 
\begin{equation}
w\in (0,0.6).  \label{weff}
\end{equation}

The most obvious candidate, a cosmological constant, which provides $w=0,$
has unappealing features: its value would be one hundred orders of magnitude
smaller than dimensionally expected; upper limits from lensing effects
barely allows for a $\Omega _{\Lambda }=0.7$ (Kochanek 1995), as would be
necessary to reconcile the amount of matter in clusters with the flatness
suggested by inflation. The next simplest possibility is perhaps to include
in the cosmic fluid a light scalar field. In fact, if the field is light
enough to vary slowly during a Hubble time, its potential energy can drive
an accelerated expansion, just like during inflation. The varying field
equation of state can then be tuned to lie in the observed range: if this is
the case, then the scalar field is sometimes denoted in the literature as \
''quintessence''. The scalar field density fraction $\Omega _{\phi }$ can be
made to decrease rapidly in the past, so as to pass easily the lensing
constraints, and to avoid discrepancies in the primordial nucleosynthesis
abundances.

Beside the acceleration argument, a light scalar field is interesting on its
own. First, it is predicted by many fundamental theories (string theory,
pseudo-Nambu-Goldstone model, Brans-Dicke theory etc.), so that it is
natural to look at its cosmological consequences (Ratra \& Peebles 1988,
Wetterich 1995, Frieman et al. 1995, Ferreira \& Joyce 1998). Second, the
presence of a scalar field may fix the standard CDM spectrum (Zlatev et al.
1998, Caldwell et al. 1998, Perrotta \& Baccigalupi 1999, Viana \& Liddle
1998). Finally, even a small amount of scalar field density may give a
detectable contribution to the standard CDM scenario, similar to what one
has in the MDM model (Ferreira \& Joyce 1998).

A scalar field, however, is expected to couple explicitly (that is, beyond
the gravitational coupling) to ordinary matter, with a strength comparable
to gravity, as put into evidence by Carroll (1998), unless some special
symmetry prevents or suppresses the coupling. Such a strong coupling would
render the scalar field interaction as strong as gravity, and would
therefore have been already detected. However, a residual coupling still
below detection cannot be excluded; moreover, if the coupling to baryons is
different from the coupling to dark matter, as proposed by Damour et al.
(1990), then even a strong coupling is indeed possible. Exactly the same
arguments hold if one supposes the quintessence field to be coupled to
gravity, rather than to matter, as investigated by Uzan (1999), Chen \&
Kamionkowsky (1999) and Baccigalupi et al. (1999). Indeed, the two models,
although physically different, are related mathematically by a conformal
transformation (Wetterich 1995, Amendola 1999a).

The non-minimal coupling of the quintessence field to ordinary matter is
therefore worth investigating, especially because the wealth of
high-precision data that is near to come allows the intriguing possibility
of detecting the coupling on the microwave background and on the present
galaxy distribution. In a previous paper (Amendola 1999b, hereinafter Paper
I) I showed that a scalar field with an exponential potential (Wetterich
1988, Ratra \& Peebles 1988) and an explicit coupling to matter may behave
as a kind of hot dark matter component, as was first shown by Ferreira \&
Joyce (1998) for zero coupling. I showed that the CMB spectrum of the model
presents acoustic peaks \ displaced from their location without coupling,
and that the galaxy power spectrum also bends in agreement to real data. In
that case, the field density amounts to at most 20\% of the critical
density, and the expansion is not accelerated. The interesting feature was
that the universe has always been in an attractor solution, independently of
the initial conditions.

In this paper I focus instead on accelerated solutions. I explore first the
general phase space of a homogeneous quintessence model with the same
exponential potential and coupling to matter as in Paper I; I will refer to
this model as coupled quintessence, or CQ. Once the phase-space attractors
have been identified, two distinct solutions are selected that allow an
accelerated epoch, and the density fluctuations on these trajectories are
studied by the use of a purposedly modified version of the code CMBFAST by
Seljak \& Zaldarriaga (1995). The linear perturbations in the uncoupled case
has been already studied by Viana \& Liddle (1998) and Caldwell et al.
(1998). As it will be shown, the coupling introduces several qualitatively
new features.

\section{Coupled scalar field model}

Consider two components, a scalar field $\phi $ and ordinary matter (e.g.,
baryons plus CDM) described by the energy-momentum tensors $T_{\mu \nu (\phi
)}$ and $T_{\mu \nu (m)}$, respectively. General covariance requires the
conservation of their sum, so that it is possible to consider a coupling
such that, for instance, 
\begin{eqnarray}
T_{\nu (\phi );\mu }^{\mu } &=&CT_{(m)}\phi _{;\mu },  \nonumber \\
T_{\nu (m);\mu }^{\mu } &=&-CT_{(m)}\phi _{;\mu }.  \label{coup1}
\end{eqnarray}
Such a coupling arises for instance in string theory, or after a conformal
transformation of Brans-Dicke theory (Wetterich 1995, Amendola 1999a). It
has also been proposed to explain 'fifth-force' experiments, since it
corresponds to a new interaction that can compete with gravity and be
material-dependent. A coupling that violates general covariance is instead
studied in Barrow \& Magueijo (1999).

The specific coupling (\ref{coup1}) is only one of the possible forms.
Non-linear couplings as $CT_{(m)}F(\phi )\phi _{;\mu }$ or more complicate
functions are also possible. Also, one can think of different couplings to
different matter species, for instance coupling the scalar field only to
dark matter and not to baryons, as proposed by Damour et al. (1990) and
Casas et al. (1992). Notice that the coupling to radiation (subscript $ 
\gamma $) vanishes, since $T_{(\gamma )}=0.$ Here I restrict myself to the
simplest possibility (\ref{coup1}), which is also the same investigated
earlier by Wetterich (1995) and is the kind of coupling that arises from
Brans-Dicke models. In fact, a field with a gravity-coupling term $\frac{1}{2 
}\xi \phi ^{2}R$ in the Lagrangian acquires, after conformal transformation,
a coupling to matter of the form (\ref{coup1}). In the limit of small
positive coupling one obtains 
\begin{equation}
C=\kappa \sqrt{\xi }.  \label{c}
\end{equation}
where $\kappa ^{2}=8\pi M_{P}^{-2}$ and $M_{P}$ is the Planck mass.
Moreover, if the Brans-Dicke Lagrangian contains a power-law potential $ 
V(\phi )\sim \phi ^{n}$, then the transformed field $\phi ^{\prime }$
acquires an exponential potential that, for small positive $\xi $, can be
written as 
\begin{equation}
V(\phi ^{\prime })\sim \exp \left[ \left( n-4\right) \kappa \sqrt{\xi }\phi
^{\prime }\right]   \label{v}
\end{equation}
(see e.g. Futamase \& Maeda 1989, Amendola et al. 1993, Amendola 1999a). The
CQ model with a linear coupling and an exponential potential that is studied
here is therefore conformally equivalent to a large class of Brans-Dicke
Lagrangians.

There are several constraints on the coupling constant $C$ along with
constraints on the mass of the scalar field particles, reviewed by Ellis et
al. (1989) and Damour (1996). The constraints arise from a variety of
observations, ranging from Cavendish-type experiments, to primordial
nucleosynthesis bounds, to stellar structure, etc. Most of them, however,
apply only if the scalar field couples universally to all matter, which is
not necessarily the case. The most stringent bound for a universal coupling,
quoted by Wetterich (1995), amounts to 
\begin{equation}
|C|<0.1M_{P}^{-1}.  \label{wetlimit}
\end{equation}
If the coupling to dark matter is different from the baryon coupling, then
the constraints on the former relaxes considerably, and becomes (Damour et
al. 1990) 
\begin{equation}
|C|<5M_{P}^{-1}.  \label{damlimit}
\end{equation}
It is also to notice that these constraints are local both in space and
time, and could be easily escaped by a time-dependent coupling constant. In
the following, therefore, $C$ is left as a free parameter.

The constraints from nucleosynthesis refer to the energy density in the
scalar component. This has to be small enough not to perturb element
production, so that at the epoch of nucleosynthesis (Wetterich 1995, Sarkar
1996, Ferreira \& Joyce 1998) 
\begin{equation}
\Omega _{\phi }(\tau _{ns})<0.1-0.2.  \label{ns}
\end{equation}
This bound is satisfied by all the interesting models.

\section{Background}

Here I derive the background equations in the flat conformal FRW metric 
\[
ds^{2}=a^{2}(-d\tau ^{2}+\delta _{ij}dx^{i}dx^{j}). 
\]
The CQ scalar field equation is 
\begin{equation}
\ddot{\phi}+2H\dot{\phi}+a^{2}U_{,\phi }=C\rho _{m}a^{2},  \label{kg}
\end{equation}
where $H=\dot{a}/a$ , and I adopt the exponential potential 
\begin{equation}
U(\phi )=Ae^{\sqrt{\frac{2}{3}}\kappa \mu \phi }.
\end{equation}
The matter (subscript $m$) and the radiation (subscript $\gamma $) equations
are 
\begin{eqnarray}
\dot{\rho}_{m}+3H\rho _{m} &=&-C\rho _{m}\dot{\phi} \\
\dot{\rho}_{\gamma }+4H\rho _{\gamma } &=&0.
\end{eqnarray}
Denoting with $\tau _{0}$ the conformal time today, let us put 
\begin{equation}
a(\tau _{0})=1,\quad \rho _{m}(\tau _{0})=\frac{3H_{0}^{2}}{8\pi }\Omega
_{m}=\rho _{m0},\quad \rho _{\gamma }(\tau _{0})=\rho _{\gamma 0},\quad \phi
(\tau _{0})=\phi _{0}.
\end{equation}
Without loss of generality, the scalar field can be rescaled by a constant
quantity, by a suitable redefinition of the potential constant $A$. We put
then $\phi _{0}=0$. This gives 
\begin{eqnarray}
\rho _{m} &=&\rho _{m0}a^{-3}e^{-C\phi },  \nonumber \\
\rho _{\gamma } &=&\rho _{\gamma 0}a^{-4}.
\end{eqnarray}
The Friedman-Einstein equation can be written as 
\begin{equation}
H^{2}=\frac{\kappa ^{2}}{3}\left( \frac{\rho _{m0}}{a}e^{-C\phi }+\frac{\rho
_{\gamma 0}}{a^{2}}+\frac{1}{2}\dot{\phi}^{2}+Ua^{2}\right) .  \label{hub}
\end{equation}

The dynamics of the CQ model has been analysed in Amendola (1999a) in the
regime in which either matter or radiation dominates. Here I generalize the
analysis to the case in which both matter and radiation are present. As we
will see, this introduces some interesting new features. Generalizing
Copeland et al. (1997) the following variables are introduced: 
\begin{equation}
x=\frac{\kappa }{H}\frac{\dot{\phi}}{\sqrt{6}},\quad y=\frac{\kappa }{H} 
\sqrt{\frac{U}{3}},\quad z=\frac{\kappa }{H}\sqrt{\frac{\rho _{\gamma }}{3},}
\label{basic}
\end{equation}
along with the independent variable $\alpha =\log a$. Notice that $x^{2}$ , $ 
y^{2}$ and $z^{2}$ give the fraction of total energy density carried by the
field kinetic energy, the field potential energy, and the radiation,
respectively, that is $\Omega _{\phi }=x^{2}+y^{2}$ and $\Omega _{\gamma
}=z^{2}$. Clearly, the matter energy density fraction is the complement to
unity of $x^{2}+y^{2}+z^{2}$. We can rewrite the Eq. (\ref{kg}) and ( 
\ref{hub}) as 
\begin{eqnarray}
x^{\prime } &=&\left( \frac{z^{\prime }}{z}-1\right) -\mu y^{2}+\beta
(1-x^{2}-y^{2}-z^{2}),  \nonumber \\
y^{\prime } &=&\mu xy+y\left( 2+\frac{z^{\prime }}{z}\right) ,  \nonumber \\
z^{\prime } &=&-\frac{z}{2}\left( 1-3x^{2}+3y^{2}-z^{2}\right) ,  \label{sys}
\end{eqnarray}
where the prime denotes here $d/d\alpha $ and where I introduced the
dimensionless constant 
\begin{equation}
\beta =\sqrt{\frac{3}{2}}\frac{C}{\kappa }\quad   \label{bec}
\end{equation}
(in Amendola 1999a $\beta $ was defined as twice the value above). The
parameters $\beta $ and $\mu $ are all we need to completely specify our
model. The constraints quoted in the previous section on $C$ become now $ 
\beta <0.025$ for the universal coupling and $\beta <1$ for the dark matter
coupling.

The system (\ref{sys}) is invariant under the change of sign of $y,z$ and of 
$\alpha $. Since it is also limited by the condition $\rho >0$ to the circle 
$x^{2}+y^{2}+z^{2}\leq 1$, we limit the analysis only to the quarter of
unitary sphere of positive $y,z$. The critical points, i.e. the points that
verify $x^{\prime }=y^{\prime }=z^{\prime }=0$, are scaling solutions, on
which the scalar field equation of state is 
\begin{equation}
w_{\phi }=\frac{2x^{2}}{x^{2}+y^{2}}=\text{const},
\end{equation}
the scalar field total energy density is $\Omega _{\phi }=x^{2}+y^{2}$, and
the scale factor is 
\begin{equation}
a\sim \tau ^{\frac{p}{1-p}}=t^{p},\quad p=\frac{2}{3w_{\text{eff}}}
\end{equation}
($t$ being the time defined by $dt=a(\tau )d\tau $). The effective equation
of state for the total cosmic fluid, $p_{tot}=(w_{\text{eff}}-1)\rho _{tot}$
has index 
\[
w_{\text{eff}}=1+x^{2}-y^{2}+z^{2}/3=1+\Omega _{\gamma }(w_{\gamma
}-1)+\Omega _{\phi }(w_{\phi }-1) 
\]
(where $w_{\gamma }=4/3$ is the radiation equation of state). As already
noticed, a value $0<w_{\phi }<0.6$ is required by observations, while $w_{ 
\text{eff}}<2/3$ is enough for acceleration.

The system (\ref{sys}) with an exponential potential has up to fifteen
critical points, of which only eight can be in the allowed region. They are
labelled by a letter that reproduces the classification given in Amendola
1999a, and a subscript that denotes whether beside the field there is a
component of matter (subscript M), radiation (R), both (RM) or neither of
the two (in which case the energy density is taken up completely by the
scalar field; no subscript in this case). The critical points are listed in
Tab. I, where $g(\beta ,\mu )\equiv 4\beta ^{2}+4\beta \mu +18.$ For every
value of the parameters $\mu ,\beta $ there is one and only one stable
critical point (attractor); one or more saddle points can also exist. More
details on the phase space dynamics in Wetterich (1995) and Amendola
(1999a). 
\[
\begin{tabular}{|c|c|c|c|c|c|c|c|}
\hline
Point & $x$ & $y$ & $z$ & $\Omega _{\phi }$ & $p$ & $w_{\text{eff}}$ & $ 
w_{\phi }$ \\ \hline
$a$ & $-\frac{\mu }{3}$ & $\left( 1-\frac{\mu ^{2}}{9}\right) ^{1/2}$ & $0$
& $1$ & $3/\mu ^{2}$ & $2\mu ^{2}/9$ & $2\mu ^{2}/9$ \\ \hline
$b_{R}$ & $-\frac{2}{\mu }$ & $\frac{\sqrt{2}}{|\mu |}$ & $\left( 1-\frac{6}{ 
\mu ^{2}}\right) ^{1/2}$ & $\frac{6}{\mu ^{2}}$ & $1/2$ & $4/3$ & $4/3$ \\ 
\hline
$b_{M}$ & $-\frac{3}{2\left( \mu +\beta \right) }$ & $\frac{\left(
g-9\right) ^{1/2}}{2|\mu +\beta |}$ & $0$ & $\frac{g}{4\left( \beta +\mu
\right) ^{2}}$ & $\frac{2}{3}\left( 1+\frac{\beta }{\mu }\right) $ & $\frac{ 
\mu }{\mu +\beta }$ & $\frac{18}{g}$ \\ \hline
$c_{R}$ & $0$ & $0$ & $1$ & $0$ & $1/2$ & $4/3$ & $-$ \\ \hline
$c_{RM}$ & $\frac{1}{2\beta }$ & $0$ & $\left( 1-\frac{3}{4\beta ^{2}} 
\right) ^{1/2}$ & $\frac{1}{4\beta ^{2}}$ & $1/2$ & $4/3$ & $2$ \\ \hline
$c_{M}$ & $\frac{2}{3}\beta $ & $0$ & $0$ & $\frac{4}{9}\beta ^{2}$ & $\frac{ 
6}{4\beta ^{2}+9}$ & $1+\frac{4\beta ^{2}}{9}$ & $2$ \\ \hline
$d$ & $-1$ & $0$ & $0$ & $1$ & $1/3$ & $2$ & $2$ \\ \hline
$e$ & $+1$ & $0$ & $0$ & $1$ & $1/3$ & $2$ & $2$ \\ \hline
\multicolumn{8}{|c|}{Tab. I. Critical points.} \\ \hline
\end{tabular}
\]

The regions of existence and stability of the critical points are summarized
in Tab. II . In the table the attention is restricted to the half plane $\mu
>0$, since there is complete symmetry under simultaneous sign exchange of $ 
\mu $ and $\beta $. I defined 
\begin{eqnarray*}
\mu _{+} &=&\frac{1}{2}\left( -\beta +\sqrt{18+\beta ^{2}}\right) , \\
\mu _{0} &=&-\beta -\frac{9}{2\beta }.
\end{eqnarray*}
Fig. 1 displays the regions of the parameter space $\mu >0$ in which the
various points are stable. 
\[
\begin{tabular}{|l|l|l|l|}
\hline
Point & Existence & Stability & Acceleration \\ \hline
$a$ & $\mu <3$ & $\mu <\mu _{+}$,$\quad \mu <\sqrt{6}$ & $\mu <\sqrt{3}$ \\ 
\hline
$b_{R}$ & $\mu >\sqrt{6}$ & $0<\mu <-4\beta $ & never \\ \hline
$b_{M}$ & $|\mu +\beta |>3/2,\mu >\mu _{0}$ & $\mu >\mu _{+,}\quad \mu
>-4\beta $ & $\mu <2\beta $ \\ \hline
$c_{R}$ & $\forall \mu ,\beta $ & unstable $\forall \mu ,\beta $ & never \\ 
\hline
$c_{RM}$ & $|\beta |>\sqrt{3}/2$ & $\mu >-4\beta $ & never \\ \hline
$c_{M}$ & $|\beta |<3/2$ & $|\beta |<\sqrt{3}/2,\mu <\mu _{0}$ & never \\ 
\hline
$d$ & $\forall \mu ,\beta $ & unstable $\forall \mu ,\beta $ & never \\ 
\hline
$e$ & $\forall \mu ,\beta $ & unstable $\forall \mu ,\beta $ & never \\ 
\hline
\multicolumn{4}{|c|}{Tab. II. Properties of the critical points.} \\ \hline
\end{tabular}
\]

There are only two critical points that admit accelerated solutions, i.e.
solutions that satisfy (\ref{weff}): the points $a$ and $b_{M}$. They differ
in several important aspects, so we study them separately.

{\bf Solutions of type} $a$.

The attractor $a$, once reached, brings to zero the matter density. To allow
for the observed matter content of the universe, we have to select the
initial conditions, if they exist, in such a way that the attractor is not
yet reached at the present time, but the expansion is already accelerated.
On the positive side, this attractor is accelerated also for small values of
the coupling constant. Before discussing in detail the solutions, let us
notice that the limit $\beta =\mu =0$ corresponds to the ordinary
cosmological constant. Suppose then we put initially the field at zero
kinetic energy ($x=0$). A trajectory acceptable from a cosmological point of
view should start deep into the radiation era ($z\approx 1$), then enter a
matter dominate era ($z\approx 0$), and finally fall into the attractor $a$,
the only one still existent for $\beta =\mu =0$, which corresponds to the $ 
\Lambda $-dominated stage. In other words, the path of a ordinary $\Lambda $
universe would be $c_{R}\rightarrow c_{M}\rightarrow a$. A similar sequence
of critical points characterizes all the trajectories discussed in the
following.

In Fig. 2 we show the 3D phase space $\left( x,y,z\right) $ of model $a$,
with $\beta =0,\mu =0.1$ and $\beta =0.5,\mu =0.1$ . As before, a
cosmological trajectory must start in the radiation era ($z\simeq 1$) and
has to provide the correct final conditions for $x,y$ and $H.$ Since the
scalar field has to start dominating only recently, it is clear that its
initial energy density deep in the radiation era has to be very small: in
the 3D phase space this means that the cosmic solutions start near $ 
(x,y,z)=(0,0,1)$ , that is, near the unstable critical point $c_{R}$. The
trajectories in Fig. 2 that fall almost vertically from top are examples of
such cosmic solutions. The attractor of the CQ model is the same as for the
uncoupled case, at $\left( x,y,z\right) =\left( -0.033,0.99,0\right) $, but
while in the ordinary quintessence case there is a saddle point $c_{M\text{ } 
}$ at $\left( x,y,z\right) =\left( 0,0,0\right) $, in CQ this moves to $ 
\left( x,y,z\right) =\left( 2\beta /3,0,0\right) $, on which $\Omega _{\phi
}=4\beta ^{2}/9$. This appears more clearly from Fig. 3, in which the trend
of $\Omega _{m},\Omega _{\phi }$ and $\Omega _{\gamma }$ is reported. The
path of this solution is then $c_{R}\rightarrow c_{M}\rightarrow a$, just
as in the pure $\Lambda $ model. For $\beta >3/2$, actually, the point $ 
c_{M} $ ceases to exist, but such high values of $\beta $ are anyway
unacceptable.

It is to be emphasized that the stage of constant and finite $\Omega _{\phi
}=4\beta ^{2}/9$ is typical of the CQ model, as it is absent in the limit of
zero coupling. Let us call this the field-matter-dominated era, or $\phi $MDE 
$.$ As we will see, this stage is responsible of most of the differences
with respect to ordinary quintessence.

The equivalence time in CQ occurs earlier than in uncoupled quintessence: 
\begin{equation}
a_{eq}=\left( \frac{\rho _{\gamma 0}}{\rho _{m0}}\right) ^{\frac{3}{3-2\beta
^{2}}}.
\end{equation}
For the acceptable values of $\beta $, however, this shift has only a minor
effect.

Finally, it is to be noticed that the attractor $a$ is independent of the
coupling $\beta .$ Then, as the universe at $z\simeq 5$ converges toward the
point $a$, the dynamics becomes independent of the coupling. As a
consequence, the cosmological probes at $z<5$ , like the type Ia supernovae
or the cluster abundance, are not efficient tools for discriminating between
ordinary quintessence and CQ.

{\bf Solutions of type} $b_{M}$.

The attractor $b_{M}$ is a solution for which the matter density and the
scalar field share a finite and {\it constant} portion of the cosmic energy.
For instance, if we put 
\begin{equation}
\beta =4.02,\quad \mu =2.68
\end{equation}
we get $\Omega _{M}=0.3$ , $\Omega _{\phi }=0.7$ and $w_{\phi }\simeq 0.14,$
well within the requested range. These values, once reached, remain
indefinitely constant. The coincidence of similar values of the energy
density in the matter and field component is therefore solved, independently
of the initial conditions. On the negative side, however, these solutions
require a large value of the coupling constant ( $\beta >\sqrt{6/5}$ to
obtain $w_{\phi }<0.6$) and are therefore at risk of running into conflict
with constraints on the coupling derived from local experiments. The
strongest objection to these solutions, however, comes from the simple fact
that they lack a matter dominated era, as we will show in a moment.

In Fig. 4 we show the phase space of model $b_{M}$, assuming $\beta
=4.02,\quad \mu =2.68$. As can be seen, the phase space is now completely
different. The attractor is at $\left( x,y,z\right) =\left(
-0.22,0.81,0\right) $. The trajectory that falls from top is again similar
to the one used in the perturbation calculations. As can be seen better in
Fig. 5, there is a transient near the saddle point $c_{RM}$ , here at $ 
\left( x,y,z\right) \simeq \left( 0.12,0,0.95\right) $. In Fig. 5 the
evolution of $\Omega _{m},\Omega _{\phi }$ and $\Omega _{\gamma }$ in two
cases is displayed, one for which the present value of $w_{\text{eff}}$
equals 0.4, for which $\beta $ and $\mu $ must be chosen as above, and the
other for $w_{\text{eff}}=0.5,$ which requires $\beta =\mu =2.37$. \ In both
cases we can see the transient $c_{RM}$ , characterized by $\Omega _{\phi
}=1/(4\beta ^{2})$ $\ $and $\Omega _{M}=1-1/(2\beta ^{2})$, followed by the
decay of the radiation component and the stabilization of the field and
matter to their final values. The path of this solution is then $ 
c_{R}\rightarrow c_{RM}\rightarrow b_{M}$, in contrast with the solutions of
the type $a$. As already remarked, such a trajectory is possible only for $ 
\beta \neq 0$. The most conspicuous features here is that the radiation
dominates until recently ($z\simeq 50$) and the matter dominated era is
absent. As is intuitive, such behavior is catastrophic for the growth of the
fluctuations: when a fluctuation mode reenters the horizon, in fact, is
suppressed first by the long RDE, \ and then by the accelerated expansion.
As a consequence, the present $\sigma _{8}$ is unacceptably small, of the
order of $10^{-4}$ for COBE normalized spectra. Unless some mechanism other
than gravitational instability powers the fluctuations, the trajectories of
type $b_{M}$ are precluded. In the following we restrict therefore our
attention to solutions of type $a$.

This concludes the analysis of the homogeneous solutions of CQ. It is to be
stressed that we considered all the possible accelerated solutions. As in
the next section we will span all the range of $\beta $ and $\mu $ that
yield cosmologically acceptable solutions of type $a$, we can consider
exhausted the analysis of CQ for exponential potentials and linear coupling.

\section{Perturbations along solutions $a$}

We now proceed to study the evolution of the perturbations in the coupled
quintessence theory. The equations of the perturbations in the synchronous
gauge have been derived and discussed at length in Paper I and will not be
repeated here. In that case it was shown that several important features of
the perturbation evolution could be derived analytically, since the
background evolution was always on one of the attractor, and thus
particularly simple. The same occurs here, at least in some cases. As
described in Paper I, all the results presented below have been obtained by
modifying the code CMBFAST of Seljak \& Zaldarriaga (1996).

The key fluctuation equation in Paper I was the evolution of the sub-horizon
perturbations in the MDE regime, the only situation in which the evolution
differs from the pure CDM case. For CQ, this regime is actually the $\phi $ 
MDE introduced above. Let us note first that along any attractor with $ 
x=x_{a}$ one has, from Eq. (\ref{basic}) 
\begin{equation}
\phi =\frac{\sqrt{6}x_{a}}{\kappa }\log a.
\end{equation}
Denoting with $\delta $ the fluctuation in the matter component, we derived
in Paper I an equation for sub-horizon modes in MDE along an attractor that
can be rewritten as follows: 
\begin{equation}
\ddot{\delta}+H\left( 1+2x_{a}\beta \right) \dot{\delta}-\frac{3}{2} 
H^{2}\Omega _{M}\delta \left( 1-\frac{4\beta ^{2}}{3}\right) =0.
\label{mdesub}
\end{equation}
As we noticed in the previous section, the solutions $a$ passes through the $ 
\phi $MDE transient attractor $c_{M}$, before the present $\phi $-dominated
epoch. Therefore, from Tab. I, we can substitute $x_{a}=2\beta /3$. It
follows that Eq. (\ref{mdesub}) has a growing solution 
\begin{equation}
\delta =Aa^{m},\quad \text{with}\quad m=1-\frac{4\beta ^{2}}{3}.  \label{sup}
\end{equation}
This shows two important facts: first, the fluctuations in $\phi $MDE are
suppressed with respect to the standard CDM behavior ($m=1$), which also
holds for the uncoupled quintessence model, for all values of $\beta $;
second, the evolution does not depend on the sign of $\beta $. B
efore the present time, at $z\simeq 5$, the trajectory deviates from the $ 
\phi $MDE solution, and $\phi $ begins to dominate. 

In Fig. 6 the behavior of three density fluctuation wavelengths, calculated
numerically with CMBFAST , is shown as a function of the scale factor $a$. I
plot $\delta /a$ to enhance the differences among the various cases. It is
possible to distinguish four distinct epochs. Let us first follow the
intermediate wavelength of Fig. 6. First, the 100 Mpc/$h$ fluctuation grow
as $a^{2}$, while it is a super-horizon mode in RDE; then, around $z\simeq
2500$ $(a\simeq .0004)$ , it reenters the horizon and freezes until the $ 
\phi $MDE begins. In uncoupled quintessence ($\beta=0$) 
the fluctuation grows now as $a$ 
, as usual; in CQ the growth is instead suppressed as expected from Eq. (\ref
{sup}). Finally, around $z\simeq 5$, $\phi $ starts to dominate, the
universe accelerates, and the fluctuation growth is definitely suppressed in
all cases. The longer wavelength follows the same phases, except that it
reenters directly in the $\phi $MDE, and therefore bypasses the freezing
stage. The shorter wavelength starts in the plot already inside the horizon,
and follows the same $\phi $MDE growth evolution of the other modes.

The same $\phi $MDE attractor solution can be used to derive the location of
the first acoustic peak on the CMB. In fact, this depends essentially on the
size of the sound horizon $r_{s}$ at decoupling (subscript $d$). We have
shown in Paper I that the following approximated law governs the size of the
sound horizon along an attractor solution 
\begin{equation}
r_{s}=r_{0}\frac{a_{d}^{x_{a}\beta }}{1+2x_{a}\beta }
\end{equation}
where $r_{0}=2(a_{d}/3)^{1/2}H_{0}^{-1}$ is the standard sound horizon.
Therefore, the multipole location of the first acoustic peak is 
\begin{eqnarray}
\ell _{peak} &\simeq &\frac{2\pi }{r_{s}H_{0}}=\ell _{0}(1+2x_{a}\beta
)a_{d}^{-x_{a}\beta } \\
&=&\ell _{0}\left( 1+\frac{4\beta ^{2}}{3}\right) a_{d}^{-2\beta ^{2}/3}
\end{eqnarray}
where $\ell _{0}\simeq 2\pi /(r_{0}H_{0})\simeq 200$ is the standard peak
location, and where the second line specializes to the $\phi $MDE attractor $ 
c_{M}$. For instance, $\beta =0.1$ gives a location $\ell _{peak}\simeq
1.06\ell _{0}$ which, at least to a first approximation, is in agreement
with the numerical results below. Notice again that the peak displacement is
always toward larger $\ell $s, regardless of the sign of $\beta $, contrary
to what happened in Paper I along the attractor $b_{M}$. Similar behavior
was found numerically by Chen \& Kamionkowsky (1999) and Baccigalupi et al.
(1999) in Brans-Dicke models. Here, substituting $\beta ^{2}=3\xi /2$, we
find that the peak shifts approximately as $(1+2\xi )a_{d}^{-\xi }.$

The solution we use as background in this section is, as anticipated, on its
way to the attractor $a$. The initial conditions will be chosen so that at
the present time (i.e., when $H^{-1}=3000$ Mpc$/h$) we have $\Omega _{m}=0.3$ 
, $\Omega _{\phi }=0.7$ and $w_{\phi }\simeq 0.$ This implies that at the
present time we should have 
\begin{equation}
x_{0}\simeq 0,\quad y_{0}=0.837,\quad z_{0}=0.092
\end{equation}
independently of $\beta $. We begin by investigating the parameter range: 
\[
\mu =0.1,\quad \beta \in \left( 0,0.15\right) 
\]
The initial conditions that produce the requested final values have been
obtained by trial and errors. Inserting the background solution in the
modified CMBFAST code, we obtain the CMB spectra reported in Fig. 7. The
other values adopted are 
\begin{equation}
h=0.7,\quad \Omega _{b}=0.04,\quad n=1.
\end{equation}
It can be noticed that the peaks move to larger multipoles, as expected.
Their amplitude is generally reduced, due both to the growth suppression
found above, and to the fact that now the COBE normalization at small $\ell $
includes the integrated Sachs-Wolfe (ISW) effect, no longer negligible in
CQ, and as a consequence the fluctuation amplitude at decoupling is reduced
(see for instance Hu \& White 1996). The ISW is also responsible for the
tilt at small multipoles. Models with $\beta >0.15$ are already ruled out by
CMB observations, while models with $\beta <0.01$ are essentially
indistinguishable from uncoupled quintessence. Values smaller than $\beta
<0.06$ produce acoustic peaks which are slightly above those for \ the
uncoupled model. The bound $\beta <0.15$ is already stronger than (\ref
{damlimit}), valid for the coupling to dark matter; the determination of the
spectrum with a precision of 5\%, within reach of the Boomerang or Maxima
experiments will constrain $\beta $ to two decimal digits, better than
current constraints to the baryon coupling (\ref{wetlimit}). The effect for $ 
\beta <0.03$, which satisfies the constraint for the universal coupling
model, will be distinguishable in the near future.

The reason for the small increase in the acoustic peak that is observed for $ 
|\beta |<0.06$ is not entirely clear. Since the numerical fluctuation growth
follows very closely the analytical prediction of Eq. (\ref{sup}) I believe
the rise is not a numerical artifact. Notice that
for small $\beta $ the $\phi $MDE transient attractor starts just past the
decoupling epoch, and thus the analytical expectations based on the $\phi $ 
MDE solution are not accurate.

The effect of changing $\mu $, within the limit $\mu <\sqrt{3}$ necessary to
have acceleration, is minimal, since the trajectories must anyway satisfy
the same final conditions. The spectrum for $\beta =0$ is therefore almost
identical to the spectrum of a pure $\Lambda $ model with $\Omega _{\Lambda
}=0.7$ . Also, as expected from the analytical expressions, the perturbative
results are almost insensitive to the sign of $\beta .$ The present analysis
therefore spans all the possible accelerated trajectories in CQ with $\Omega
_{\phi }=0.7$ that are cosmologically acceptable.

In Fig. 8 I report the power spectra $\Delta ^{2}(k)=k^{3}P(k)/(2\pi ^{2})$
normalized to COBE, compared to the data compiled and corrected for redshift
and non-linear distortions by Peacock \& Dodds (1994). It can be seen that
the slope of the spectrum is in rough agreement with what is observed (the
largest discrepancy is for the four smallest scale data points, where
non-linearity and redshift distortions are more difficult to correct); a
more precise comparison depends on the assumption that the bias between
galaxies and dark matter is scale-independent, and on other variables which
are not of interest here, like $h$.

The matter fluctuation variance in 8Mpc$/h$ cells $\sigma _{8}$ for a $ 
\Omega _{m}=0.3$ universe should be around unity to fit the cluster
abundance (White, Efstathiou \& Frenk 1993, Viana \& Liddle 1996, Girardi et
al. 1998). Wang \& Steinhardt (1998) find, for a constant-$w$ \ model, a
general expression for $\sigma _{8},$ that corresponds to $\sigma _{8}\in
(0.85,1.25)$ at the 95\% c.l., adopting our cosmological parameters. It is
found that this is satisfied by $|\beta |\leq 0.1$ , so that this can be
taken as the upper limit on $|\beta |$ (see Fig. 9). As found analytically,
the suppression of $\sigma _{8}$ with respect to COBE-normalized standard
CDM is due to the growth suppression in MDE. Another factor is that now the
COBE normalization includes the integrated Sachs-Wolfe effect, no longer
negligible. The rise in the CMB spectrum that we observed for small $\beta $
induces via the COBE normalization a similar small rise in $\sigma _{8}$ for
the same values, as can be seen in Fig. 9.

A fit to $\sigma _{8}(\beta )$ gives 
\begin{equation}
\sigma _{8}(\beta )=\sigma _{8(0)}10^{\left( 4.5\beta \right) ^{1.5}-\left(
6.4\beta \right) ^{2}}
\end{equation}
where $\sigma _{8(0)}$ refers to uncoupled quintessence, and contains all
the dependence on the other cosmological parameters, as well as on the exact
COBE normalization scheme (I used here the Bunn \& White (1997)
normalization implemented in the original CMBFAST code).

\section{Conclusion}

Soon cosmology will benefit of high-precision data that will allow
unprecedented accuracy in testing fundamental theories. Quintessence models
add to the battery of cosmological parameters at least two entries, one
describing the potential of the field and another its coupling to the rest
of the world. So far, the coupling was arbitrarily put to zero, although we
know no symmetry condition that implies so. In this paper we let the
coupling be non zero, and investigated systematically all the possible
trajectories in a flat space that lead to accelerated expansion at the
present with $\Omega _{\phi }=0.7$.

The results for the homogeneous theory are applicable also to all
Brans-Dicke models with a power law potential, since there is a direct
correspondence between our constants $\beta $ and $\mu $ and the Brans-Dicke
coupling constant $\xi $ and the potential exponent $n$; from Eqs. (\ref{c}, 
\ref{v},\ref{bec}) we have in fact
\begin{eqnarray}
\beta ^{2} &=&\frac{3}{2}\xi ,  \nonumber \\
\mu ^{2} &=&\frac{3}{2}(n-4)^{2}\xi .
\end{eqnarray}
For the fluctuations, the transformation that brings one from the
fluctuation quantities in the Jordan frame ( the frame in which the
field is coupled to gravity) to those in the Einstein frame (in which the
field is coupled to the matter) is more complicated, and a complete
treatment is still to be published. However, in the limit in which the
fluctuations of the scalar field are small with respect to the fluctuations
in the other components the fluctuation fields are conformally invariant;
since the CQ field is almost homogeneous due to its light mass, this
condition is verified for most of the universe history. As a consequence, it
is conjectured that also the perturbative CQ results apply to Brans-Dicke
models. The verification of this conjecture is left to future work.

The main feature of the CQ model is the existence of a phase intermediate
between the radiation era and the accelerated era, that we denoted $\phi $ 
MDE. During this era the fluctuations grow less than in an uncoupled model.
The $\phi $MDE has three effects on the CMB: the spectrum at low multipoles
is tilted, due to the ISW effect; the acoustic peaks are shifted to higher
multipoles, due to the change in the sound horizon; and their amplitude is
changed in a non-trivial way. On the power spectrum at the present, the main
effect is the reduction of $\sigma _{8}$ for large couplings and a very
minor enhancement for small coupling. \ \ \ 

We found that the potential slope is not efficiently constrained by
observations, essentially because the $\phi $MDE is independent of $\mu $.
The coupling $\beta $ is on the contrary constrained already by the present
data, and is expected to be much more so in the near future, by at least an
order of magnitude. From CMB and $\sigma _{8}$ data we can derive the bound 
\begin{equation}
|\beta |<0.1
\end{equation}
which is stronger than the dark matter constraint of Damour et al. (1990)
and not very far from the universal constraint of Wetterich (1995).

\section*{Acknowledgments}

I am indebted to Carlo Baccigalupi, Francesca Perrotta and Michael Joyce for
useful discussions on the topic.



\newpage
\begin{figure*}
\epsfysize 8in
\epsfbox[-10 4 492 700]{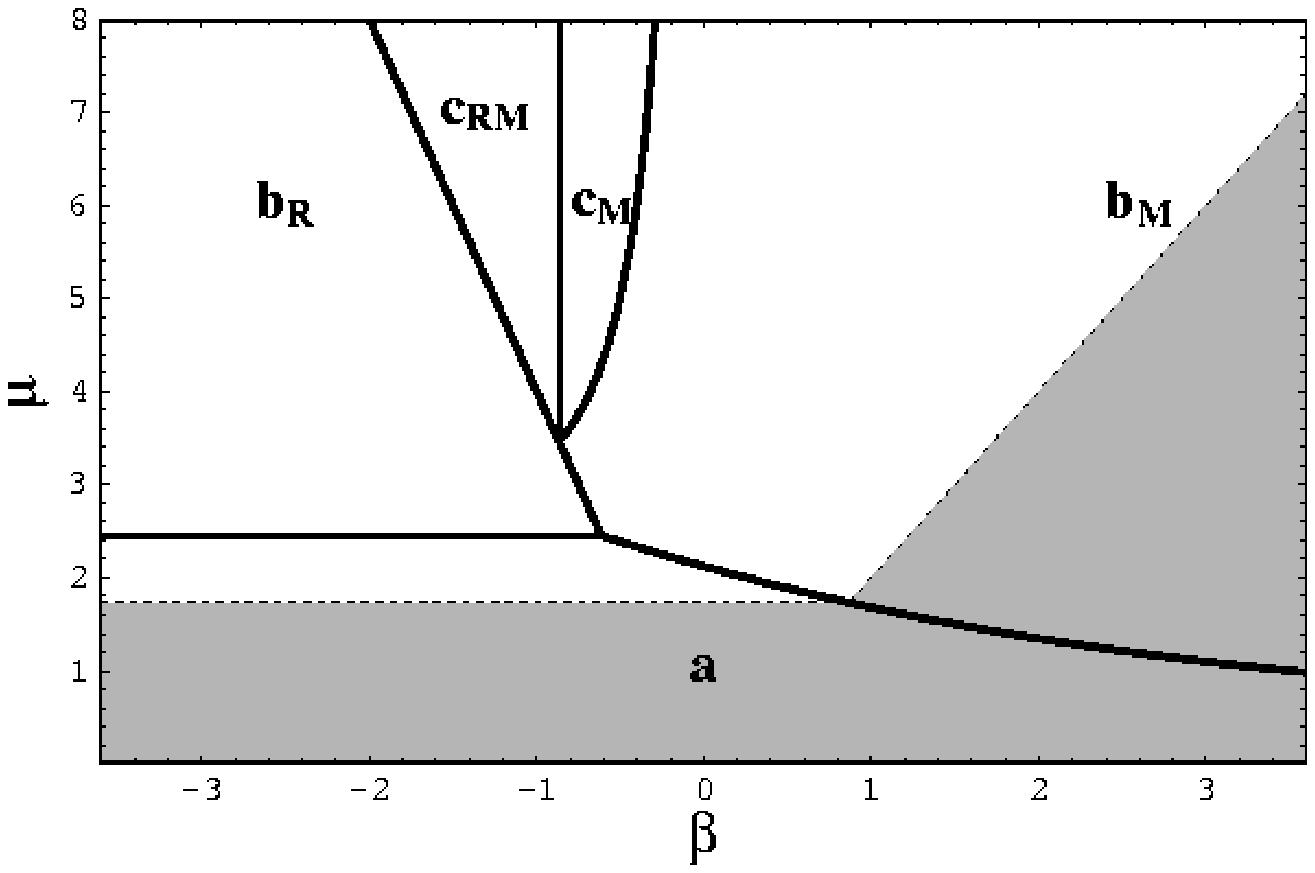}
\caption{The figure shows the parameter space of the model. Each
region is labelled by the critical point that is stable in that region. The
shaded area indicates the values for which the attractor is accelerated.}
\end{figure*}
\newpage
\begin{figure*}
\epsfysize 8in
\epsfbox[-10 4 492 700]{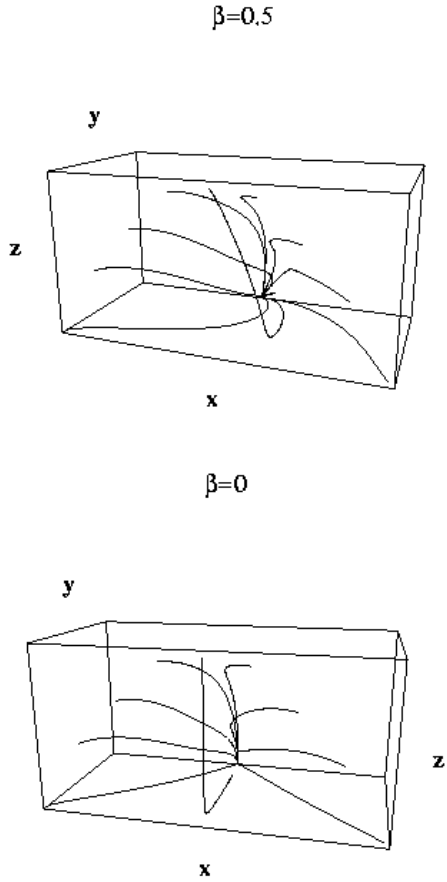}
\caption{ CQ phase space for values that lie in the $a$ region, $\mu
=0.1 $ and $\beta $ as indicated. The attractor $a$ is the same as in the
uncoupled model, but for $\beta \neq 0$ there is a saddle for a non-zero
value of the scalar field density. The phase spaces for the values of $\beta 
$ investigated in this paper are qualitatively similar to the $\beta =0.5$
case. The trajectory that falls almost vertically from top is similar to the
background solution effectively employed.}
\end{figure*}

\newpage
\begin{figure*}
\epsfysize 8in
\epsfbox[-10 4 492 700]{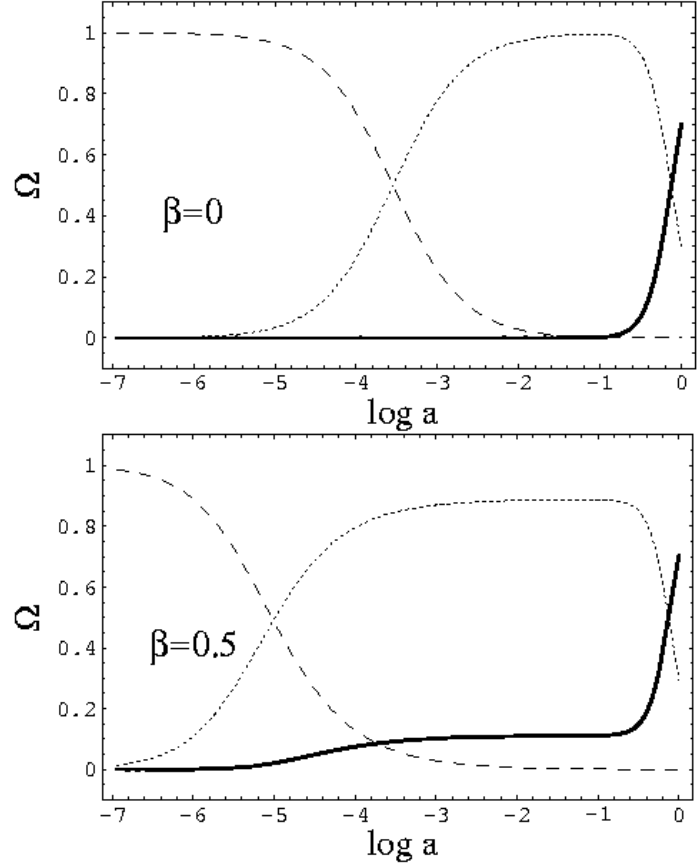}
\caption{Behavior of $\Omega _{M}$ (dotted line), $\Omega _{R}$ (dashed
line) and $\Omega _{\phi }$ (thick line) as a function of $\log (a)$ for $ 
\mu =0.1$ and $\beta $ as indicated. Notice that for CQ there is the
transient regime $\phi $MDE in which both the matter and the scalar field
energy density are non-vanishing. Notice also that in this case, and for all
values of $\beta \neq 0$, the matter-radiation equivalence occurs earlier
than in the uncoupled model. For the small values of $\beta $ used for the
perturbation calculations, however, this is a small effect. }
\end{figure*}
\newpage
\begin{figure*}
\epsfysize 8in
\epsfbox[-10 4 492 700]{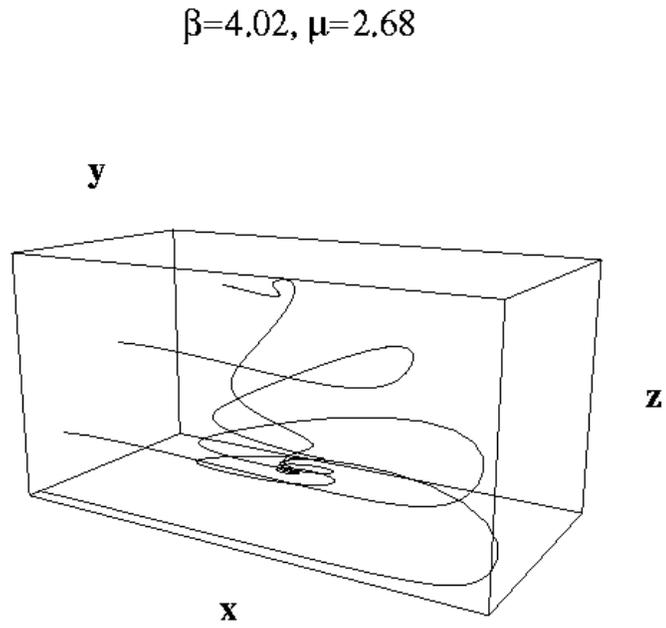}
\caption{ Phase space of CQ for values that lie in the $b_{M}$ region.
There is a saddle $c_{RM}$ at $\left( x,y,z\right) \simeq \left(
0.12,0,0.95\right) $ that attracts the trajectory that falls from top,
similar to the one adopted in the perturbation calculation.
 }
\end{figure*}
\newpage
\begin{figure*}
\epsfysize 8in
\epsfbox[-10 4 492 700]{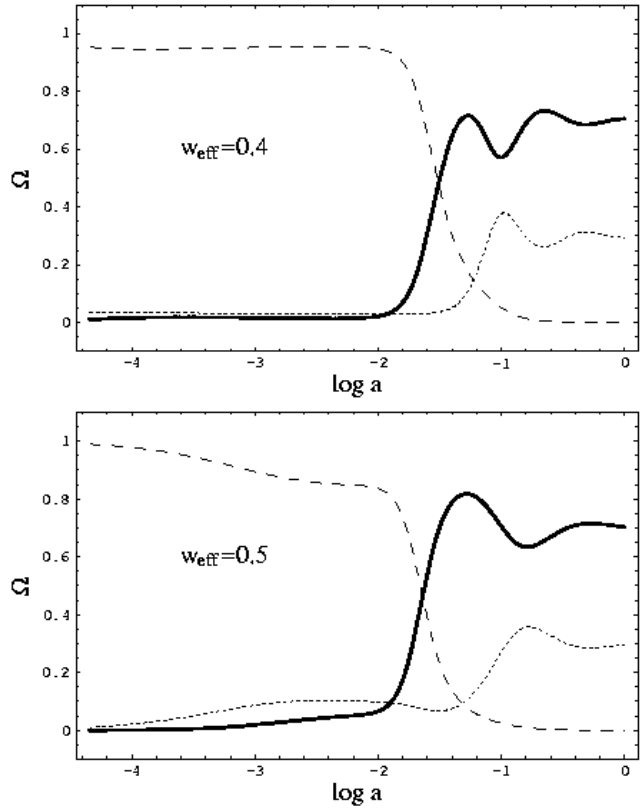}
\caption{ Behavior of $\Omega _{M}$ (dotted line), $\Omega _{R}$ (dashed
line) and $\Omega _{\phi }$ (thick line) as a function of $\log (a)$ for \
the same parameters as in Fig. 4 (label $w_{\text{eff}}=0.4$) and for $\beta
=\mu =2.37$ (label $w_{\text{eff}}=0.5$). Notice the transient epoch in
which radiation and field share the total energy density (saddle $c_{RM}$).
}
\end{figure*}
\newpage
\begin{figure*}
\epsfysize 8in
\epsfbox[-10 4 492 700]{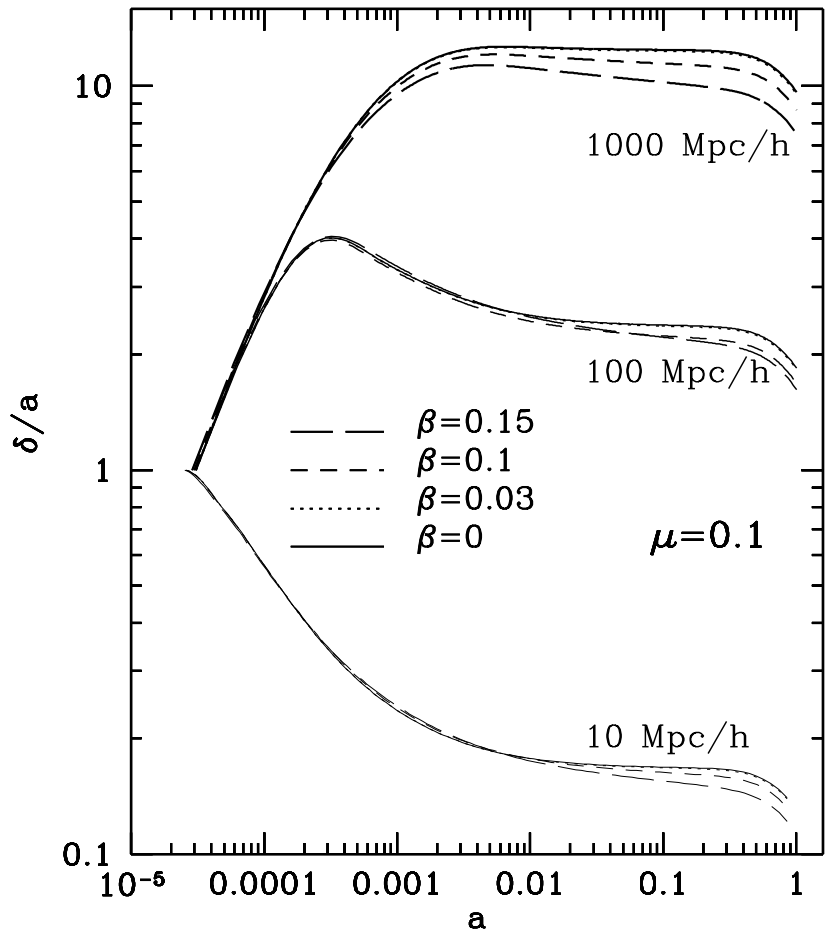}
\caption{Evolution of the matter density contrast $\delta /a$ for three
wavelengths, 2$\pi /k=\lambda =$1000 Mpc$/h$ , 100 Mpc$/h$ and 10 Mpc$/h$,
for $\mu =0.1$ and various values of $\beta $. Notice the growth suppression
for $\beta \neq 0$. }
\end{figure*}
\newpage
\begin{figure*}
\epsfysize 8in
\epsfbox[-10 4 492 700]{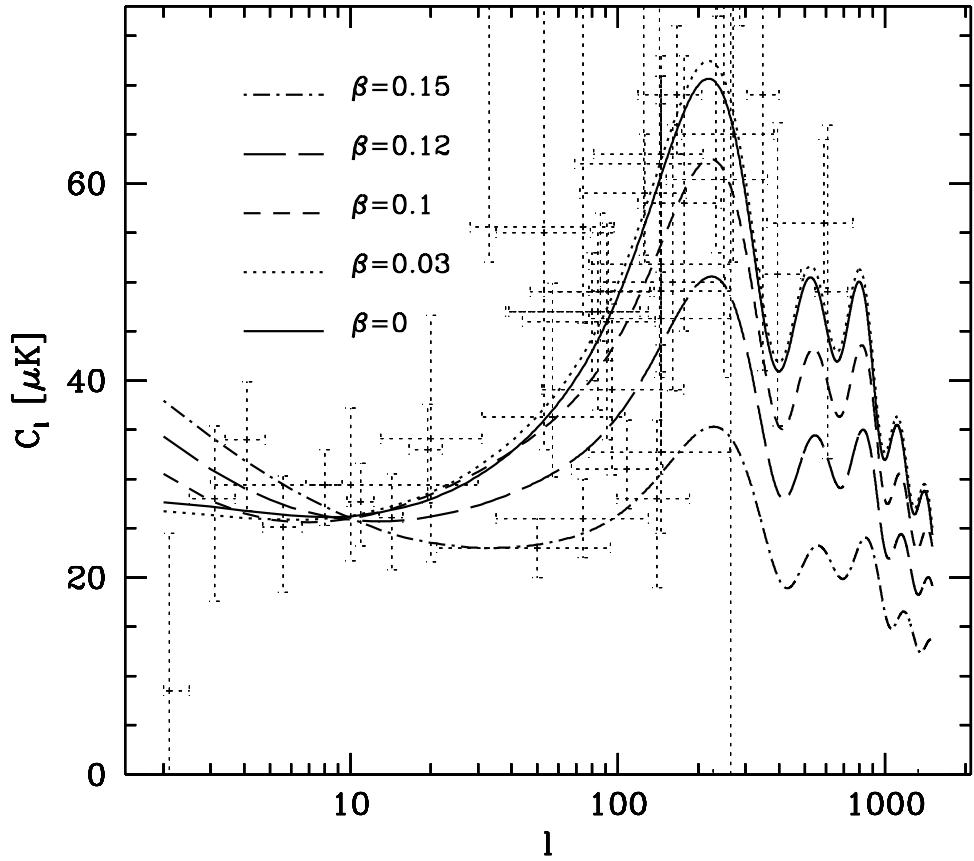}
\caption{Multipole spectra $C_{\ell }$ (actually we plot $\left[ \ell
\left( \ell +1\right) C_{\ell }/2\pi \right] ^{1/2}\mu $K, as customary) for
the solutions $a$. The data are a selection from Tegmark's home page
(http://www.sns.ias.edu/ \symbol{126}max). }
\end{figure*}

\begin{figure*}
\epsfysize 8in
\epsfbox[-10 4 492 700]{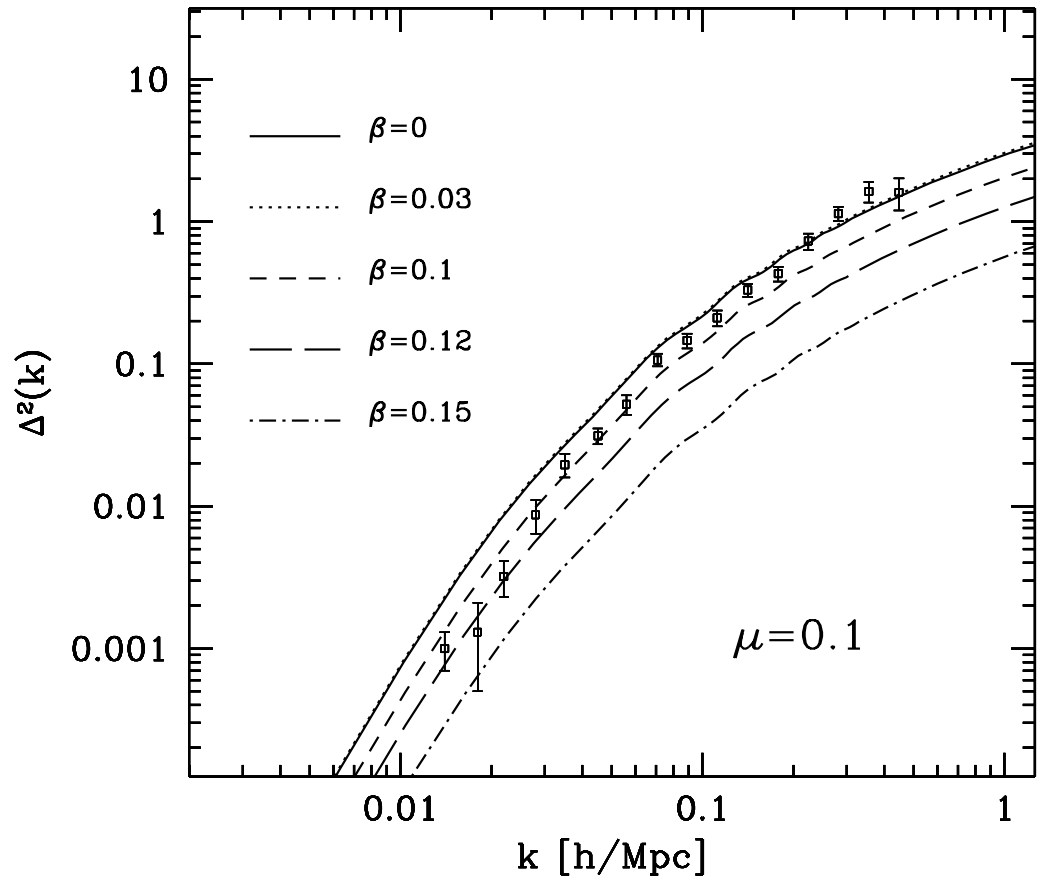}
\caption{ Power spectrum $\Delta ^{2}(k)=k^{3}P(k)/(2\pi ^{2})$ for the
same solutions of type $a$ of Fig. \ 7. The real galaxy data are from
Peacock \& Dodd (1994).}
\end{figure*}
\newpage
\begin{figure*}
\epsfysize 8in
\epsfbox[-10 4 492 700]{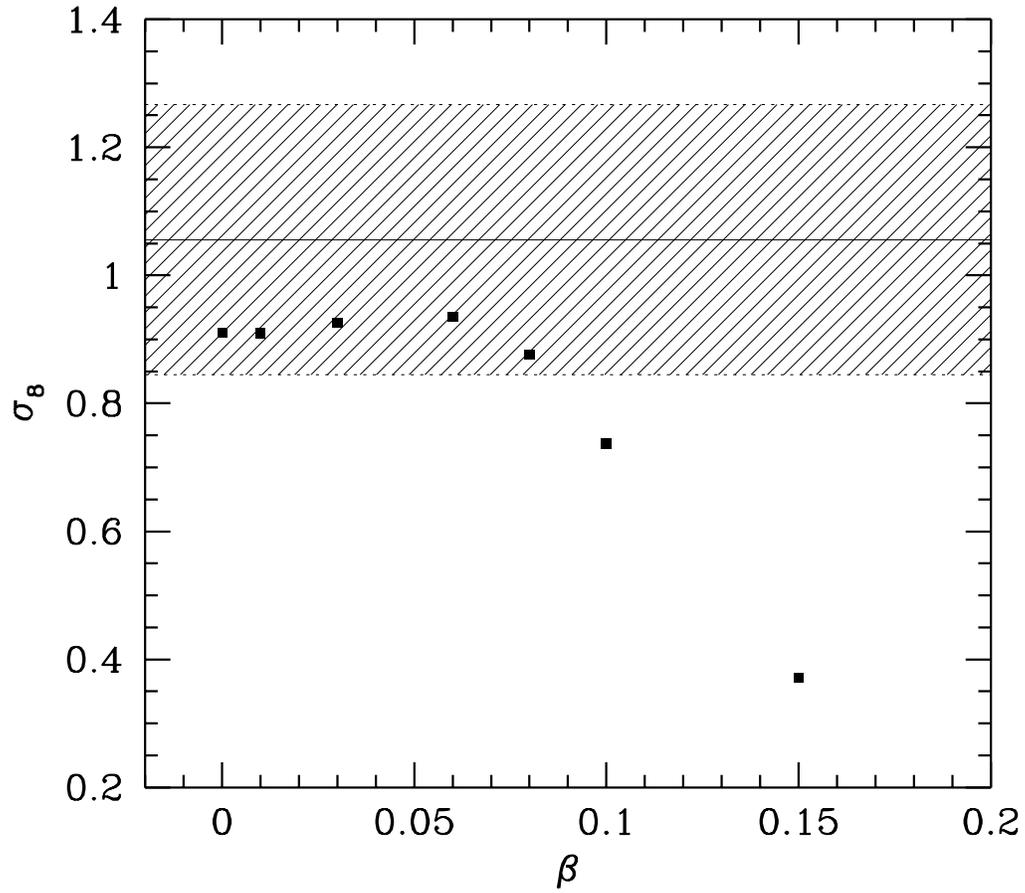}
\caption{ The variance $\sigma _{8}$ versus $\beta .$ The shaded area is
the 95\% c.l. region that matches the cluster abundance (Wang \& Steinhardt
1998).}
\end{figure*}

\end{document}